\documentclass[preprint]{aastex}

\usepackage{euscript,epsfig,amsmath,amssymb,amsfonts,latexsym}

\newcommand{\be}[1]{\begin{equation}\label{#1}}
\newcommand{\ee}{\end{equation}}
\newcommand{\ba}[1]{\begin{eqnarray}\label{#1}}
\newcommand{\ea}{\end{eqnarray}}
\newcommand{\rf}[1]{(\ref{#1})}
\newcommand{\nn}{\nonumber}

\shorttitle{First-order cosmological perturbations} \shortauthors{M. Eingorn}

\begin{document}

\title{First-order Cosmological Perturbations Engendered by Point-like Masses}

\author{Maxim Eingorn}
\affil{North Carolina Central University, CREST and NASA Research Centers\\ Fayetteville st. 1801, Durham, North Carolina 27707, U.S.A.\\}
\email{maxim.eingorn@gmail.com}

\begin{abstract}
In the framework of the concordance cosmological model the first-order scalar and vector perturbations of the homogeneous background are derived in the weak
gravitational field limit without any supplementary approximations. The sources of these perturbations (inhomogeneities) are presented in the discrete form of a
system of separate point-like gravitating masses. The found expressions for the metric corrections are valid at all (sub-horizon and super-horizon) scales and
converge at all points except at locations of the sources. The average values of these metric corrections are zero (thus, first-order backreaction effects are
absent). Both the Minkowski background limit and the Newtonian cosmological approximation are reached under certain well-defined conditions. An important feature
of the velocity-independent part of the scalar perturbation is revealed: up to an additive constant this part represents a sum of Yukawa potentials produced by
inhomogeneities with the same finite time-dependent Yukawa interaction range. The suggested connection between this range and the homogeneity scale is briefly
discussed along with other possible physical implications.
\end{abstract}


\

\section{INTRODUCTION}

The concordance cosmological model fits well the contemporary observations \citep{WMAP,Planck13,Planck15,BAO}. This model assumes that the Universe is filled with
dominant portions of cold dark matter (CDM) and dark energy, represented by the cosmological constant and assuring the acceleration of the global expansion, as
well as a comparatively small portion of standard baryonic matter and a negligible portion of radiation. According to the cosmological principle, on large enough
scales the Universe is treated as being homogeneous and isotropic, so the corresponding background Friedmann-Lema\^{\i}tre-Robertson-Walker (FLRW) metric is
appropriate for its description. On the contrary, on sufficiently small scales the Universe is highly inhomogeneous (separate galaxies, galaxy groups and clusters
stare us in the face). The thorough theoretical study of the structure formation starting from primordial fluctuations at the earliest evolution stages with the
subsequent comparison of the predictions with the cosmic microwave background and other observational data may be recognized as one of the major subjects of
modern cosmology.

The structure growth is usually investigated by means of two main distinct approaches, namely, the relativistic perturbation theory (see, e.g.,
\citet{Bardeen,Ruth,Rubakov}) and $N$-body simulations generally based on the Newtonian cosmological approximation (see, e.g., \citet{Springel,Dolag,Chisari}).
Roughly speaking, the area of the first approach may be characterized by the keywords ``early Universe; linearity; large scales'', while the area of the second
one may be defined by the keywords of the opposite meaning: ``late Universe; nonlinearity; small scales''. Both approaches make great progress in describing the
inhomogeneous world within the limits of their applicability. Nevertheless, the linear cosmological perturbations theory certainly fails in describing nonlinear
dynamics at small distances, while Newtonian simulations do not take into account relativistic effects becoming non-negligible at large distances
\citep{Wald,AdamekRuth,Adamek2,Bruni}. In this connection, in the latter case certain effort is required for extracting relativistic features from the large-scale
Newtonian description \citep{Chisari,gauge,Hahn}.

Until now there was no developed unified scheme, which would be valid for arbitrary (sub-horizon and super-horizon) scales and treat the non-uniform matter
density in the non-perturbative way, thereby incorporating its linear and nonlinear deviations from the average. This acute problem is addressed and successfully
resolved in the present paper by construction of such a self-consistent and indispensable scheme, which promises to be very useful in the precision cosmology era.

A couple of previous similar attempts deserves mentioning. First, the generalization of the well-known nonrelativistic post-Minkowski formalism \citep{Landau} to
the cosmological case in the form of the relativistic post-Friedmann formalism, which would be valid on all scales and include the full nonlinearity of Newtonian
gravity at small distances, has been made in the recent paper by \citet{Bruni}, but the authors resorted to expansion of the metric in powers of the parameter
$1/c$ (the inverse speed of light). Second, the formalism for relativistic $N$-body simulations in the weak field regime, suitable for cosmological applications,
has been developed by \citet{AdamekRuth}, but the authors gave different orders of smallness to the metric corrections and to their spatial derivatives (a similar
``dictionary'' can be found in \citep{Wald,Adamek2}).

The current paper also relies on the weak gravitational field regime: deviations of the metric coefficients
%as well as the energy-momentum tensor components
from their background (average) values are considered as first-order quantities, while the second order is completely disregarded. However, there are no
additional assumptions: in the spirit of relativity, spatial and temporal derivatives are treated on an equal footing, and no dictionary giving them different
orders of smallness is used (in contrast to \citet{Wald,AdamekRuth,Adamek2}); expansion into series with respect to the ratio $1/c$ is not used as well (in
contrast to \citet{Bruni}); there is no artificial mixing of first- and second-order contributions; the sub- or super-horizon regions are not singled out, and the
derived formulas for the metric corrections are suitable at all scales.

The desired formalism is elaborated below within discrete cosmology \citep{EZcosm1,EZcosm2,EKZ2,Gibbons,Ellis}, based on the well-grounded idea of presenting
nonrelativistic matter in the form of separate point-like particles with the corresponding energy-momentum tensor \citep{Landau}. At sub-horizon scales,
implementation of this idea leads to nonrelativistic gravitational potentials and Newtonian equations of motion against the homogeneous background
\citep{Peebles}, which are commonly used in modern $N$-body simulations. Now the explicit expressions for potentials, applicable at super-horizon scales as well,
will be made available.

The paper is structured in the following way. Section 2 is entirely devoted to solving the linearized Einstein equations for the first-order scalar and vector
cosmological perturbations of the homogeneous background. In Section 3, the arresting attention properties of the derived solutions, including their asymptotic
behavior, are analyzed and their role in addressing different related physical challenges is indicated. The main results are summarized laconically in the
Conclusion.

\

\section{DISCRETE PICTURE OF COSMOLOGICAL PERTURBATIONS}

\setcounter{equation}{0}

\subsection{Equations}

The unperturbed FLRW metric, describing the Universe being homogeneous and isotropic on the average, reads:
\be{1} ds^2=a^2\left(d\eta^2-\delta_{\alpha\beta}dx^{\alpha}dx^{\beta}\right),\quad \alpha,\beta=1,2,3\, , \ee
where $a(\eta)$ is the scale factor; $\eta$ is the conformal time;  $x^{\alpha}$, $\alpha=1,2,3$, stand for the comoving coordinates, and it is supposed for
simplicity that the spatial curvature is zero (the generalization to the case of non-flat spatial geometry is briefly analyzed below). The corresponding Friedmann
equations in the framework of the pure $\Lambda$CDM model read:
\be{2} \frac{3{\mathcal H}^2}{a^2}=\kappa\overline{\varepsilon} + \Lambda\ee
and
\be{3} \frac{2{\mathcal H}'+{\mathcal H}^2}{a^2}=\Lambda\, , \ee
where ${\mathcal H}\equiv a'/a\equiv (da/d\eta)/a$; the prime denotes the derivative with respect to $\eta$; $\kappa\equiv 8\pi G_N/c^4$ ($c$ is the speed of
light and $G_N$ is the Newtonian gravitational constant); $\varepsilon$ represents the energy density of the nonrelativistic pressureless matter; the overline
indicates the average value, and $\Lambda$ is the cosmological constant. Domination of cold matter and $\Lambda$ is at the center of attention, so contributions
of radiation, relativistic cosmic neutrinos or any warm component are negligible.

Following the analysis of the first-order cosmological perturbations by \citet{Bardeen,Ruth,Rubakov}, let us fix the Poisson gauge and consider the respective
metric
\be{4} ds^2=a^2\left[\left(1 + 2\Phi\right)d\eta^2 + 2B_{\alpha}dx^{\alpha}d\eta - \left(1- 2\Phi\right)\delta_{\alpha\beta}dx^{\alpha}dx^{\beta}\right]\, , \ee
where the function $\Phi(\eta,{\bf r})$ and the spatial vector ${\bf B}(\eta,{\bf r})\equiv(B_1,B_2,B_3)$ describe the scalar and vector perturbations,
respectively. It is assumed that there is no anisotropic stress acting as a source of the difference between perturbations of the metric coefficients $g_{00}$ and
$g_{\alpha\beta}$, $\alpha=\beta$. Therefore, both these metric corrections are equated to the same expression $2a^2\Phi$ from the very beginning. Tensor
perturbations are not taken into account because their source is also of the order beyond the adopted accuracy. The first-order tensor perturbations are
associated with gravitational waves, freely propagating against the FLRW background. Their propagation is governed by the well-known equation uncoupled from the
equations for $\Phi$ and ${\bf B}$ (see, e.g., Eq.~(11) in \citep{Hwang}). Here attention is concentrated solely on the perturbations with non-negligible sources,
and cosmological gravitational waves are not investigated. Similarly to, e.g., \citet{AdamekRuth,Adamek2}, it is demanded that
\be{5} \nabla {\bf B}\equiv \delta^{\alpha\beta}\frac{\partial B_{\alpha}}{\partial x^{\beta}}=0\, .\ee
Then the Einstein equations
\be{6} G_i^k=\kappa T_i^k+\Lambda\delta_i^k,\quad i,k=0,1,2,3\, ,\ee
where $G_i^k$ and $T_i^k$ denote the mixed components of the Einstein tensor and the matter energy-momentum tensor, respectively, take the following form after
linearization:
\be{7} G_0^0=\kappa T_0^0+\Lambda \quad\Rightarrow\quad \triangle\Phi-3{\mathcal H}(\Phi'+{\mathcal H}\Phi) = \frac{1}{2}\kappa a^2{\delta T}_{0}^0\, , \ee
\be{8} G_{\alpha}^0=\kappa T_{\alpha}^0 \quad\Rightarrow\quad \frac{1}{4}\triangle{B}_{\alpha}+\frac{\partial}{\partial x^{\alpha}}(\Phi'+{\mathcal
H}\Phi)=\frac{1}{2}\kappa a^2{\delta T}_{\alpha}^0\, , \ee
\ba{9} G_{\alpha}^{\beta}=\kappa T_{\alpha}^{\beta}+\Lambda\delta_{\alpha}^{\beta} \quad\Rightarrow\quad \Phi''+3{\mathcal H}\Phi'+\left(2{\mathcal H}'+{\mathcal
H}^2\right)\Phi=0\, , \ea
\ba{10} \left(\frac{\partial B_{\alpha}}{\partial x^{\beta}}+\frac{\partial B_{\beta}}{\partial x^{\alpha}}\right)'+2{\mathcal H}\left(\frac{\partial
B_{\alpha}}{\partial x^{\beta}}+\frac{\partial B_{\beta}}{\partial x^{\alpha}}\right)=0\, . \ea
Here $\triangle\equiv\delta^{\alpha\beta}\cfrac{\partial^2}{\partial x^{\alpha}\partial x^{\beta}}$ stands for the Laplace operator in the comoving coordinates;
$T_i^k= \overline{T}_i^k+\delta T_i^k$, where $\overline{T}_0^0=\overline\varepsilon$ is the only nonzero average mixed component. In the spirit of the
particle-particle method of $N$-body simulations, the matter constituent of the Universe may be presented in the form of separate point-like massive particles.
Then the deviations $\delta T_i^k$ from the average values $\overline{T}_i^k$ can be easily determined with the help of the well-known general expression for the
matter energy-momentum tensor contravariant components \citep{Landau,Chisari,EZcosm1,EZcosm2}:
\be{11} T^{ik}=\sum\limits_n\frac{m_nc^2}{\sqrt{-g}}\frac{dx_n^i}{d\eta}\frac{dx_n^k}{d\eta}\frac{d\eta}{ds_n}\delta({\bf r}-{\bf r}_n)\, ,\ee
where $g\equiv\mathrm{det}(g_{ik})$. This expression corresponds to a system of gravitating masses $m_n$ with the comoving radius-vectors ${\bf r}_n(\eta)$ and
$4$-velocities $u^i_n\equiv dx^i_n/ds_n$, $i=0,1,2,3$. Their rest mass density $\rho(\eta,{\bf r})$ in the comoving coordinates reads:
\be{12} \rho=\sum\limits_nm_n\delta({\bf r}-{\bf r}_n)=\sum\limits_n\rho_n,\quad \rho_n\equiv m_n\delta({\bf r}-{\bf r}_n)\, .\ee

Introducing the comoving peculiar velocities $\tilde v^{\alpha}_n\equiv dx^{\alpha}_n/d\eta$, $\alpha=1,2,3$, and treating them as importing the first order of
smallness in the right-hand side (rhs) of the linearized Einstein equations \rf{7} and \rf{8}, one finds out that within the adopted accuracy
\be{13} \delta T_0^0\equiv T_0^0-\overline{T}_0^0=\frac{c^2}{a^3}\delta\rho+\frac{3\overline\rho c^2}{a^3}\Phi,\quad \delta\rho\equiv\rho-\overline\rho\, , \ee
and
\be{14} \delta T_{\alpha}^0=-\frac{c^2}{a^3}\sum\limits_nm_n\delta({\bf r}-{\bf r}_n)\tilde v^{\alpha}_n+\frac{\overline\rho
c^2}{a^3}B_{\alpha}=-\frac{c^2}{a^3}\sum\limits_n\rho_n\tilde v^{\alpha}_n+\frac{\overline\rho c^2}{a^3}B_{\alpha}\, ,\ee
respectively. In addition, $\delta T_{\alpha}^{\beta}=0$. In particular, in full agreement with, e.g., \citet{AdamekRuth}, the first-order anisotropic stress is
considered as vanishing for nonrelativistic matter. Consequently, the rhs of both Eqs.~\rf{9} and \rf{10} are zero. In the formula \rf{13} the arising term
$\sim\,\rho\Phi$ is replaced by the term $\sim\,\overline{\rho}\Phi$ since the product $\delta\rho\Phi$ imports the second order of smallness and therefore should
be dropped (the inequality $|\delta\rho|\gg|\delta\rho\Phi|$ certainly holds true at all scales \citep{Chisari}). The same reasoning applies to the term
$\sim\,\rho B_{\alpha}$ arising in the formula \rf{14}. The average rest mass density $\overline{\rho}$ is related to the average energy density
$\overline\varepsilon$ by means of an evident equality: $\overline\varepsilon=\overline\rho c^2/a^3$.

It is crucial that throughout the present paper, similarly to \citet{Chisari,EZcosm1,EZcosm2,AdamekRuth,Adamek2}, the rest mass density $\rho$ is treated in the
non-perturbative way: fulfilment of the inequality $|\delta\rho|\ll\overline\rho$ is not demanded. For instance, the intragalactic medium and dark matter halos
are characterized by the values of $\rho$ being much higher than $\overline\rho$. Thus, nonlinearity with respect to the deviation of $\rho$ from its average
value $\overline\rho$ at small scales is fully taken into consideration here. The sole requirement lies in the following: this deviation $\delta\rho$ as a source
of the metric correction $\Phi$ must secure smallness of this correction, i.e. the inequality $|\Phi|\ll1$. As regards the Einstein tensor components, their
nonlinear deviation from the corresponding background values is also unforbidden.
%(so that both sides of Eqs.~\rf{7}-\rf{10} are of the same order)
For instance, the term $\sim\triangle\Phi$ in the expression for $G_0^0$ predominates at small scales not only over the other terms in this expression, but also
over the background value $3\mathcal{H}^2/a^2$ of $G_0^0$. Nevertheless, as distinct from \citet{AdamekRuth,Adamek2}, those terms which are nonlinear with respect
to the metric corrections are neglected in the expressions for the Einstein tensor components. This concerns, e.g., the product $\Phi\triangle\Phi$ which is
neglected in comparison with $\triangle\Phi$, in complete agreement with the initial well-grounded assumption $|\Phi|\ll1$.

Substitution of the expressions \rf{13} and \rf{14} into Eqs.~\rf{7} and \rf{8}, respectively, gives
\be{15} \triangle\Phi-3{\mathcal H}(\Phi'+{\mathcal H}\Phi)-\frac{3\kappa\overline\rho c^2}{2a}\Phi = \frac{\kappa c^2}{2a}\delta\rho\, ,\ee
\be{16} \frac{1}{4}\triangle{\bf B}+\nabla(\Phi'+{\mathcal H}\Phi)-\frac{\kappa\overline\rho c^2}{2a}{\bf B}=-\frac{\kappa c^2}{2a}\sum\limits_nm_n\delta({\bf
r}-{\bf r}_n)\tilde {\bf v}_n=-\frac{\kappa c^2}{2a}\sum\limits_n\rho_n\tilde {\bf v}_n\, ,\ee
where $\tilde{\bf v}_n(\eta)\equiv d{\bf r}_n/d\eta\equiv(\tilde{v}_n^1,\tilde{v}_n^2,\tilde{v}_n^3)$. With the help of
%the well-known useful auxiliary equalities $-\triangle(1/r)=\nabla\left({\bf r}/r^3\right)=4\pi\delta({\bf r})$ and
the continuity equation
\be{17} \rho'_n+\nabla(\rho_n\tilde{\bf v}_n)=0\, ,\ee
which is satisfied identically for any $n$-th particle, it is not difficult to split the vector $\sum\limits_n\rho_n\tilde {\bf v}_n$ into~its grad- and curl-
parts:
\be{18} \sum\limits_n\rho_n\tilde {\bf v}_n=\nabla\Xi+\left(\sum\limits_n\rho_n\tilde {\bf v}_n-\nabla\Xi\right)\, ,\ee
where
\be{19} \Xi=\frac{1}{4\pi}\sum\limits_nm_n\frac{({\bf r}-{\bf r}_n)\tilde {\bf v}_n}{|{\bf r}-{\bf r}_n|^3}
%=-\frac{1}{8\pi}\triangle\left[\sum\limits_nm_n\frac{({\bf r}-{\bf r}_n)\tilde {\bf v}_n}{|{\bf r}-{\bf r}_n|}\right]
\, .\ee
Really, this function satisfies the Poisson equation
\be{20} \triangle\Xi=\nabla\sum\limits_n\rho_n\tilde {\bf v}_n=-\sum\limits_n\rho'_n\, ,\ee
and this fact can be easily checked alternatively with the help of the Fourier transform:
\be{21} -k^2\hat\Xi=i{\bf k}\sum\limits_n\hat\rho_n\tilde {\bf v}_n,\quad k\equiv|{\bf k}|\, ,\ee
where
\be{22} \hat\Xi(\eta,{\bf k})\equiv \int\Xi(\eta,{\bf r})\exp(-i{\bf kr})d{\bf r}\, ,\ee
\be{23} \hat\rho_n(\eta,{\bf k})\equiv \int\rho_n(\eta,{\bf r})\exp(-i{\bf kr})d{\bf r}=m_n\int\delta({\bf r}-{\bf r}_n)\exp(-i{\bf kr})d{\bf r}=m_n\exp(-i{\bf
k}{\bf r}_n)\, .\ee
In other words, one can demonstrate that the function $\hat\Xi(\eta,{\bf k})$ derived from \rf{22} after the substitution of \rf{19} satisfies Eq.~\rf{21} and,
consequently, reads:
\be{24} \hat\Xi=-\frac{i}{k^2}\sum\limits_nm_n({\bf k}\tilde{\bf v}_n)\exp(-i{\bf k}{\bf r}_n)\, .\ee

Taking into account \rf{18}, from \rf{16} one gets
\be{25} \Phi'+{\mathcal H}\Phi=-\frac{\kappa c^2}{2a}\Xi
%+f(\eta)
%=-\frac{\kappa c^2}{8\pi a}\sum\limits_nm_n\frac{({\bf r}-{\bf r}_n)\tilde {\bf v}_n}{|{\bf r}-{\bf r}_n|^3}+f(\eta)\, ,
\ee
%
%where $f(\eta)$ is some space-independent function,
and
\be{26} \frac{1}{4}\triangle{\bf B}-\frac{\kappa\overline\rho c^2}{2a}{\bf B}=-\frac{\kappa c^2}{2a}\left(\sum\limits_n\rho_n\tilde{\bf v}_n-\nabla\Xi\right)
%=-\frac{\kappa c^2}{2a}\sum\limits_n\rho_n\tilde{\bf v}_n-\frac{\kappa c^2}{16\pi a}
%\nabla\triangle\left[\sum\limits_nm_n\frac{({\bf r}-{\bf r}_n)\tilde {\bf v}_n}{|{\bf r}-{\bf r}_n|}\right]
\, .\ee
Substitution of \rf{25} into \rf{15} gives
\be{27} \triangle\Phi-\frac{3\kappa\overline\rho c^2}{2a}\Phi = \frac{\kappa c^2}{2a}\delta\rho-\frac{3\kappa c^2{\mathcal H}}{2a}\Xi
%+3{\mathcal H}f(\eta)
\, .\ee
Thus, Eqs.~\rf{26} and \rf{27} are derived for the vector and scalar perturbations, respectively. Below their solutions are found and the fulfilment of
Eqs.~\rf{5}, \rf{9}, \rf{10} and \rf{25} is verified.

\

\subsection{Solutions}

In the Fourier space Eq.~\rf{26} takes the following form:
\be{28} -\frac{k^2}{4}\hat{\bf B}-\frac{\kappa\overline\rho c^2}{2a}\hat{\bf B}=-\frac{\kappa c^2}{2a}\left(\sum\limits_n\hat\rho_n\tilde{\bf v}_n-i{\bf
k}\hat\Xi\right) \, .\ee
Substituting \rf{23} and \rf{24} into \rf{28}, one immediately obtains
\be{29} \hat{\bf B}=\frac{2\kappa c^2}{a}\left(k^2+\frac{2\kappa\overline\rho c^2}{a}\right)^{-1}\sum\limits_nm_n\exp(-i{\bf k}{\bf r}_n)\left(\tilde{\bf
v}_n-\frac{({\bf k}\tilde{\bf v}_n)}{k^2}{\bf k}\right)\, .\ee

The condition \rf{5} is evidently satisfied since ${\bf k}\hat{\bf B}=0$. It is also not difficult to verify that Eq.~\rf{10} is fulfilled within the adopted
accuracy. In fact, it is enough to show that
\be{30} \hat{\bf B}'+2{\mathcal H}\hat{\bf B}=0\, .\ee
For this purpose let us write down the spacetime interval for the $n$-th particle
\be{31} ds_n=a\left[1+2\Phi+2B_{\alpha}\tilde v_n^{\alpha}-(1-2\Phi)\delta_{\alpha\beta}\tilde v_n^{\alpha}\tilde v_n^{\beta}\right]^{1/2}d\eta\ee
and the corresponding Lagrange equations of motion
\be{32} \left[a({\bf B}|_{{\bf r}={\bf r}_n}-\tilde{\bf v}_n)\right]'\ =\ a\nabla\Phi|_{{\bf r}={\bf r}_n}\, ,\ee
where the contributions of the considered particle itself to ${\bf B}$ and $\Phi$ are excluded as usual, so divergences are absent (in other words, the particle
moves in the gravitational field produced by the other particles). Since Eq.~\rf{32} and its consequences will be used exclusively in linearized Einstein
equations (e.g., Eq.~\rf{10}), all terms being nonlinear with respect to $\tilde{\bf v}_n$, ${\bf B}$ and $\Phi$ have been dropped in order to avoid exceeding the
adopted accuracy.
%(this is the reason of using the sign ``$\longleftrightarrow$" instead of ``$=$")
Multiplying \rf{32} by $\rho_n$ and summing up, one gets
\be{33} \rho(a{\bf B})'-\sum\limits_n\rho_n(a\tilde{\bf v}_n)'\ =\ a\rho\nabla\Phi\, .\ee
Further, in the terms containing ${\bf B}$ and $\Phi$ the rest mass density $\rho$ should be replaced by its average value $\overline\rho$ as discussed before.
Consequently,
\be{34} \sum\limits_n\rho_n(a\tilde{\bf v}_n)'\ =\ -a\overline\rho\nabla\Phi+\overline\rho(a{\bf B})'\, .\ee
Of course, in order to study dynamics of the $N$-body system one should include the removed nonlinear terms $-a\delta\rho\nabla\Phi$ and $\delta\rho(a{\bf B})'$.
However, for analyzing the Einstein equations for the first-order cosmological perturbations it is apparently enough to keep only linear terms in the equations of
motion. In the Fourier space Eq.~\rf{34} reads:
\be{35} \sum\limits_n\hat\rho_n(a\tilde{\bf v}_n)'=\sum\limits_nm_n\exp(-i{\bf k}{\bf r}_n)(a\tilde{\bf v}_n)'\ =\ -a\overline\rho\cdot i{\bf
k}\hat\Phi+\overline\rho(a\hat{\bf B})'\, .\ee
Now, expressing $\hat{\bf B}'$ from \rf{29} with the help of \rf{35} and substituting the result into \rf{30}, one arrives at the identity.

Finally, the vector perturbation ${\bf B}$ can be determined by multiplying \rf{29} by $\exp(i{\bf kr})/(2\pi)^3$ and integrating over ${\bf k}$. Cumbersome, but
straightforward calculation gives
\ba{36} {\bf B}&=& \frac{\kappa c^2}{8\pi a}\sum\limits_{n}\left[\frac{m_n\tilde{\bf v}_n}{|{\bf r}-{\bf r}_n|}\cdot
\frac{(3+2\sqrt{3}q_n+4q_n^2)\exp(-2q_n/\sqrt{3})-3}{q_n^2}\right.\nn\\
&+&\left.\frac{m_n[\tilde{\bf v}_n({\bf r}-{\bf r}_n)]}{|{\bf r}-{\bf r}_n|^3}({\bf r}-{\bf
r}_n)\cdot\frac{9-(9+6\sqrt{3}q_n+4q_n^2)\exp(-2q_n/\sqrt{3})}{q_n^2}\right]\, ,\ea
where the following convenient spatial vector is introduced:
\be{37} {\bf q}_n(\eta,{\bf r}) \equiv \sqrt{\frac{3\kappa\overline\rho c^2}{2a}}({\bf r}-{\bf r}_n),\quad q_n\equiv|{\bf q}_n|\, . \ee

Let us now switch over to Eq.~\rf{27}. In the Fourier space this equation takes the following form:
\be{38} -k^2\hat\Phi-\frac{3\kappa\overline\rho c^2}{2a}\hat\Phi = \frac{\kappa c^2}{2a}\sum\limits_n\hat\rho_n-\frac{\kappa\overline\rho
c^2}{2a}(2\pi)^3\delta({\bf k})-\frac{3\kappa c^2{\mathcal H}}{2a}\hat\Xi\, ,\ee
where the well-known presentation $(2\pi)^3\delta({\bf k})=\int\exp(-i{\bf kr})d{\bf r}$ is taken into account. Substituting \rf{23} and \rf{24} into \rf{38}, one
immediately obtains
\be{39} \hat\Phi =-\frac{\kappa c^2}{2a}\left(k^2+\frac{3\kappa\overline\rho c^2}{2a}\right)^{-1}\left[\sum\limits_nm_n\exp(-i{\bf k}{\bf r}_n)\left(1+3i{\mathcal
H}\frac{({\bf k}\tilde{\bf v}_n)}{k^2}\right)-\overline\rho(2\pi)^3\delta({\bf k})\right]\, .\ee

With the help of \rf{35} one can show that both Eqs.~\rf{9} and \rf{25} are fulfilled within the adopted accuracy. The scalar perturbation $\Phi$ itself can be
determined by multiplying \rf{39} by $\exp(i{\bf kr})/(2\pi)^3$ and integrating over ${\bf k}$. This Fourier inversion gives
\ba{40} \Phi&=&\frac{1}{3}-\frac{\kappa c^2}{8\pi a}\sum\limits_n\frac{m_n}{|{\bf r}-{\bf r}_n|}\exp(-q_n)\nn\\
&+&\frac{3\kappa c^2}{8\pi a}{\mathcal H}\sum\limits_n \frac{m_n[\tilde{\bf v}_n({\bf r}-{\bf r}_n)]}{|{\bf r}-{\bf r}_n|}\cdot\frac{1-(1+q_n)\exp(-q_n)}{q_n^2}\,
.\ea
Thus, the explicit analytical expressions \rf{36} and \rf{40} for the first-order vector and scalar cosmological perturbations, respectively, are determined for
the first time. Let us accentuate the irrefutable fact that the dictionary-based approach compels us to say goodbye to all hopes of finding analytical solutions.
Really, let us momentarily return to Eq.~\rf{27}. In \citep{AdamekRuth} the order $O(\epsilon)$ is assigned to $\Phi$ while every spatial derivative is treated as
importing the order $O(\epsilon^{-1/2})$, where $\epsilon$ is some small parameter. Consequently, $\triangle\Phi\sim O(1)$ while $[3\kappa\overline\rho
c^2/(2a)]\Phi\sim O(\epsilon)$. Concerning the rhs of Eq.~\rf{27}, following the same assignment, $\delta\rho\sim O(1)$, but $\Xi\sim O(\epsilon)$ (see
Eq.~\rf{25}). Thus, in the $O(1)$-approximation the second terms in both sides of Eq.~\rf{27} are missing, and this equation reduces to the Poisson equation of
Newtonian gravity, being inapplicable for large enough distances. \citet{AdamekRuth} overcome this difficulty by including the terms of the orders $O(1)$ and
$O(\epsilon)$ simultaneously in the same equations (i.e. by mixing their own $O(1)$- and $O(\epsilon)$-approximations). Then the term $[3\kappa\overline\rho
c^2/(2a)]\Phi$ is reinstated, but so do the nonlinear terms like $\Phi\triangle\Phi\sim O(\epsilon)$. This substantially complicates the problem leaving room for
numerical computation only. As shown in the current paper, it is mathematically logical to keep both considered terms $\triangle\Phi$ and $[3\kappa\overline\rho
c^2/(2a)]\Phi$ (linear with respect to $\Phi$), dropping all nonlinear ones such as $\Phi\triangle\Phi$. The resulting Helmholtz equation \rf{27} is not
restricted to sufficiently small distances and actually covers the whole space. Therefore, the desired formalism for cosmological $N$-body simulations is
developed here shedding hardly any blood: linearized Einstein equations are not complicated by extra nonlinear terms and admit exact analytical solutions
describing contributions of every single massive particle to the total inhomogeneous gravitational field.

In what follows the noteworthy features and advantages of the expressions \rf{36} and \rf{40} are briefly discussed.

\

\section{MENU OF PROPERTIES, BENEFITS, AND BONUSES}

\setcounter{equation}{0}

\subsection{Minkowski Background Limit}

After obtaining the solutions \rf{36} and \rf{40} of the system of Eqs.~\rf{7}-\rf{10} for the first-order cosmological perturbations, it is absolutely necessary
to study their asymptotic behavior. First of all, let us consider the Minkowski background limit:
\be{c1} a\rightarrow\mathrm{const} \quad\Rightarrow\quad \mathcal{H}\rightarrow0;\quad \overline\rho\rightarrow0\quad\Rightarrow\quad q_n\rightarrow0\, .\ee
Then instead of \rf{36} and \rf{40} one has, respectively,
\ba{c2} {\bf B}&\rightarrow&\frac{\kappa c^2}{4\pi a}\sum\limits_{n}\left[\frac{m_n\tilde{\bf v}_n}{|{\bf r}-{\bf r}_n|}+\frac{m_n[\tilde{\bf v}_n({\bf r}-{\bf
r}_n)]}{|{\bf r}-{\bf r}_n|^3}({\bf r}-{\bf r}_n)\right]\nn\\
&=&\frac{G_N}{2c^2}\sum\limits_{n}\frac{m_n}{|{\bf R}-{\bf R}_n|}\left[4\tilde{\bf v}_n+\frac{4[\tilde{\bf v}_n({\bf R}-{\bf R}_n)]}{|{\bf R}-{\bf
R}_n|}\frac{{\bf R}-{\bf R}_n}{|{\bf R}-{\bf R}_n|}\right]\, ,\ea
where the physical radius-vectors ${\bf R}\equiv a{\bf r}$, ${\bf R}_n\equiv a{\bf r}_n$ are introduced, and
\be{c3} \Phi\rightarrow-\frac{\kappa c^2}{8\pi a}\sum\limits_n\frac{m_n}{|{\bf r}-{\bf r}_n|}=-\frac{G_N}{c^2}\sum\limits_n\frac{m_n}{|{\bf R}-{\bf R}_n|}\, ,\ee
where the constant $1/3$ has been dropped since it originates in \rf{27} exclusively from the terms containing $\overline\rho$. Let us compare the asymptotic
expressions \rf{c2} and \rf{c3} with the corresponding corrections to Minkowski spacetime metric coefficients from the paragraph 106 of the textbook by
\citet{Landau}. This paragraph is devoted particularly to the weak gravitational field generated by a system of nonrelativistic point-like particles perturbing
flat spacetime geometry. Multiplying \rf{c3} by $2$, one arrives at the result which exactly coincides with the first term in the expression (106.13) in
\citep{Landau} for the metric correction $h_{00}$, as it certainly should be. In addition, the term containing $\tilde{\bf v}_n$ disappears from \rf{40} in the
considered limit (in view of the factor $\mathcal{H}$), and at the same time there is no term that is linear with respect to $\tilde{\bf v}_n$ in (106.13)
\citep{Landau}. This fact also serves as confirmation of coincidence.

As regards \rf{c2}, the only difference between this expression and (106.15) in \citep{Landau} is that the integers $4;\, 4$ are replaced by $7;\, 1$ respectively
(one should keep in mind that the comoving peculiar velocities $\tilde{\bf v}_n$ defined with respect to the conformal time $\eta$ as $d{\bf r}_n/d\eta$ are
related to those defined with respect to the synchronous time $t$ as ${\bf v}_n\equiv d{\bf r}_n/dt$ by means of an evident equality: $cdt=ad\eta\ \Rightarrow\
\tilde{\bf v}_n=a{\bf v}_n/c$). Apparently, this difference in integers represents none other than a result of different gauge conditions here and in
\citep{Landau}. Indeed, the condition \rf{5} is not demanded and, of course, does not hold true in \citep{Landau}. Correspondence between \rf{c2} and (106.15)
\citep{Landau} lies in the fact that the sum of these integers is the same: $4+4=7+1$. One can show that it equals $8$ for the other appropriate gauge choices as
well. Therefore, \rf{c2} exactly coincides with the purely vector part of (106.15) \citep{Landau}, as one can easily see by finding curl of both these
expressions.

\

\subsection{Newtonian Approximation and Homogeneity Scale}

Now let us switch over to the Newtonian cosmological approximation: $q_n\ll1$, i.e. $|{\bf r}-{\bf r}_n|\ll\sqrt{2a/(3\kappa\overline\rho c^2)}$, and peculiar
motion as a source of the gravitational field is completely ignored \citep{Chisari}, so the summands directly proportional to the velocities $\tilde{\bf v}_n$ are
omitted. Then only the scalar perturbation $\Phi$ survives in the same form \rf{c3}, where the constant $1/3$ has been dropped for the other reason: only the
gravitational potential gradient enters into equations of motion describing dynamics of the considered system of gravitating masses. These equations for any
$j$-th particle follow directly from \rf{31} and take the form
\be{c4} \ddot {\bf R}_j -\frac{\ddot a}{a} {\bf R}_j=-G_N\sum_{n\neq j}\frac{m_n\left({\bf R}_j-{\bf R}_n\right)}{\left|{\bf R}_j-{\bf R}_n\right|^3} \ee
in the physical coordinates $X^{\beta}=ax^{\beta}$, $\beta=1,2,3$, being in accordance with the corresponding equations in the papers by
\citet{Springel,Dolag,Labini,Warren,Vahe,Ellis} devoted to cosmological simulations. Here dots denote the derivatives with respect to $t$.

Let us consider two important questions. First, what are the applicability bounds for the above-mentioned inequality, which may be rewritten in the form $|{\bf
R}-{\bf R}_n|\ll\sqrt{2a^3/(3\kappa\overline\rho c^2)}$? In order to answer, one should simply calculate the rhs of this inequality:
\be{c5} \lambda\equiv\sqrt{\frac{2a^3}{3\kappa\overline\rho c^2}}=\sqrt{\frac{2c^2}{9H_0^2\Omega_M}\left(\frac{a}{a_0}\right)^3},\quad
\Omega_M\equiv\frac{\kappa\overline\rho c^4}{3H_0^2a_0^3}\, , \ee
where $a_0$ and $H_0$ are the current values of the scale factor $a$ and the Hubble parameter $H\equiv \dot a/a\equiv(da/dt)/a=c\mathcal{H}/a$, respectively.
According to \citet{Planck13,Planck15}, $H_0\approx68\, \mathrm{km\,s^{-1}\,Mpc^{-1}}$ and $\Omega_M\approx0.31$. Therefore, the current value of $\lambda$ is
$\lambda_0\approx3700\, \mathrm{Mpc}\approx 12\, \mathrm{Gly}$. It is very interesting that this Yukawa interaction range and the sizes of the largest known
cosmic structures \citep{LQG,wall,ring} are of the same order, thereby hinting at the opportunity to resolve the formidable challenge lying in the fact that their
sizes essentially exceed the previously reported epoch-independent scale of homogeneity $\sim\, 370\, \mathrm{Mpc}$ \citep{Yadav}. The authors arrived at this
underestimate by comparing the deviation of the fractal dimension, characterizing the distribution of matter, from $3$ (dimensionality of space) to its
statistical dispersion. Along with fractal analysis, their approach relies on the weak clustering limit and cosmological simulations driving $512^3$ particles in
a cube with the edge $\sim\, 1.5\, \mathrm{Gpc}$. Incidentally, this edge is less than half $\lambda_0$, and from the very beginning of such a volume-restricted
simulation it is difficult to expect any definite and reliable indications of structuring in bigger volumes. Now, if one associates the scale of homogeneity with
$\lambda$ instead, then the cosmological principle, asserting that the Universe is homogeneous and isotropic when viewed at a sufficiently large scale, is saved
and reinstated when this typical averaging scale is greater than $\lambda$. The proposed association does not mean that the homogeneity scale is equated exactly
to $\lambda$ but rather describes $\lambda$ as an approximate upper bound to the cosmic structure size, and the homogeneity scale as a distance exceeding
$\lambda$ in a few times while remaining of the same order. It is remarkable that this reasoning is actually confirmed by \citet{LiLin}. The authors defined the
scale of homogeneity as a distance at which the correlation dimension is within $1\,\%$ of $3$ (and, consequently, equals $2.97$) and fixed an upper bound to such
a distance $\sim\, 3\lambda_0$. The dependence $\lambda\sim a^{3/2}$ is noteworthy as well: the earlier the evolution stage, the smaller the scale of homogeneity.
Naturally, this is closely related to the hierarchical clustering process.

The second important question is: what are the applicability bounds for peculiar motion ignoring? In order to answer, one can consider the ratio of the third term
in \rf{40} to the second one. For a single gravitating mass $m_1$ momentarily located at the origin of coordinates (${\bf r}_1=0$) with the velocity $\tilde{\bf
v}_1$ collinear to ${\bf r}$ (for ensuring the maximum value of the scalar product $\tilde{\bf v}_1{\bf r}=\tilde v_1r$, where $\tilde v_1\equiv|\tilde{\bf
v}_1|$) this ratio amounts (up to a sign) to $3\mathcal{H}\tilde v_1r/2=3Hav_1R/\left(2c^2\right)$, where $v_1\equiv|{\bf v}_1|=c\tilde v_1/a$, $R\equiv|{\bf
R}|=ar$ and $q_1\ll1$ as before.
%Therefore, the contributions of the considered particle to the second and third terms in \rf{40} are equal to each other at the distance $R_{*}=2c^2/(3Hav_1)$.
Actually the product $av_1$ is none other than the absolute value of the particle's physical peculiar velocity. For example, with the help of the today's typical
values $(250\div500)\, \mathrm{km\,s^{-1}}$ and the inequality $R\ll3700\, \mathrm{Mpc}$ one finds that the considered ratio is much less than
$(1\div2)\times10^{-3}$. Exactly the same estimate can be made for the ratio of derivatives of the considered terms with respect to ${\bf r}$. This means that at
the scales under consideration, the gravitational force originating from the second term in the gravitational potential \rf{40}, which does not contain particle
velocities, is much stronger than that coming from the third one, which contains them.

Thus, the Newtonian cosmological approximation may be used when $|{\bf R}-{\bf R}_n|\ll\lambda$. Otherwise, at the scales comparable or greater than $\lambda$,
one should use the complete expressions for the metric corrections obtained in the previous section. In particular, the derived Yukawa-type potentials should be
used instead of the Newtonian ones in order to study formation and evolution of the largest structures in the Universe. It is necessary to understand that the
elaborated formalism results in Newtonian behavior of the considered physical system at sufficiently small distances without any relativistic corrections. Hence,
the accuracy of the developed theory is limited in this region by the standard Newtonian approach. However, the predicted Yukawa behavior at greater distances may
be considered a relativistic effect since it follows directly from Einstein equations of General Relativity.

It should be emphasized that despite the presence of those terms in \rf{36} and \rf{40}, which do not contain exponential functions, the influence of any particle
on the motion of its neighbours does drop exponentially when the distance increases. Really, with the help of \rf{30} the equations of motion \rf{32} may be
rewritten in the form
\be{new32} \left(a\tilde{\bf v}_n\right)'\ =\ -a\left(\nabla\Phi|_{{\bf r}={\bf r}_n}+\mathcal{H}{\bf B}|_{{\bf r}={\bf r}_n}\right)\, ,\ee
and this peculiar acceleration of a given particle, caused by all other gravitating masses, contains solely terms with exponential functions. Indeed, the direct
substitution of \rf{36} and \rf{40} into the rhs of \rf{new32} demonstrates that all terms without exponential functions exactly cancel each other. This fact
confirms the revealed Yukawa nature of universal gravitation. In addition, it indicates that for sufficiently small values of $a$ and nonzero separation distances
between particles they almost do not interact gravitationally (all terms containing exponential functions may be dropped under such conditions), so the system
behaves as a perfect gas undergoing the global expansion. It is also interesting that the physical screening length $\sqrt{3}\lambda/2$ from \rf{36} is less than
the counterpart $\lambda$ from \rf{40} meaning that vector modes diminish with distance faster than scalar modes. The equations of motion \rf{new32} are ready to
be used in a new generation of cosmological simulation codes (see, however, the discussion of coordinate transformations below, indicating possibilities of
reinterpretation of Newtonian simulations from a relativistic perspective). It would be quite reasonable to confront the outputs of relativistic simulations with
those of various Newtonian predecessors, thereby discriminating between them and the proposed Yukawa modification, especially regarding predictions of
peculiarities of hugest gravitationally bound objects in the Universe.

One more important detail consists in the fact that $\lambda$ does not coincide with the Hubble radius $c/H$, in contrast to the Yukawa interaction range proposed
by \citet{Signore} in order to limit gravitational effects of a particle outside its causal sphere. Really, in terms of the Hubble parameter $H$ and the
deceleration parameter $q\equiv-\ddot a/\left(aH^2\right)$,
\be{c6} \frac{1}{\lambda^2}=\frac{3H^2}{c^2}(1+q)=-\frac{3\dot H}{c^2},\quad \lambda=\frac{1}{\sqrt{3(1+q)}}\frac{c}{H}\, ,\ee
following directly from \rf{c5} and the Friedmann equations \rf{2} and \rf{3}, which may be rewritten in the form
\be{c7} \frac{3H^2}{c^2}=\frac{\kappa\overline\rho c^2}{a^3} + \Lambda\ee
and
\be{c8} \frac{H^2}{c^2}(1-2q)=\frac{3H^2+2\dot H}{c^2}=\Lambda\, , \ee
respectively. One obtains from \rf{c6} that $\lambda=c/H$ in the unique moment of time when $q=-2/3$. According to Eqs.~\rf{c7} and \rf{c8},
$2\Lambda/7=\kappa\overline\rho c^2/a^3$ at this moment, or $2\Omega_{\Lambda}/7=\Omega_M(a_0/a)^3$, where $\Omega_{\Lambda}\equiv\Lambda
c^2/\left(3H_0^2\right)\approx0.69$ \citep{Planck13,Planck15}. Hence, $\lambda=c/H$ in the near future when $a/a_0\approx1.16$. Before this moment $\lambda<c/H$,
while afterwards the opposite inequality takes place.

Likewise $\lambda$ does not coincide with a shielding length introduced by \citet{Hahn}. The authors resorted to the dominant growing mode in the framework of the
linear relativistic perturbation theory (see their Eq.~(15), which is actually a predetermined approximate solution but, nevertheless, serves as an assumed
starting point) and presented $\Phi$ in the standard form of a product of a function of time and a function of spatial coordinates. This allowed expressing
$3\mathcal{H}\left(\Phi'+{\mathcal H}\Phi\right)$ as $l^{-2}\Phi$, where $l$ is a certain time-dependent parameter, and then, after substitution into the
linearized Einstein equation $G_0^0=\kappa T_0^0+\Lambda$, declaring $l$ to be a shielding length. It should be mentioned that the same shielding mechanism may be
also discerned in the preceding paper by \citet{Ruslan}, where continuous matter sources are in the attention focus instead of discrete ones investigated here. In
this connection, it makes sense to confront in brief the approaches by \citet{Hahn} and \citet{Ruslan}. First, at the same level of linear energy-momentum
fluctuations, the velocity-dependent term introduced by \citet{Ruslan} in the equation for $\Phi$ (see, e.g., their Eq.~(16)) can be also easily reduced to
$l^{-2}\Phi$ for the considered growing mode. Of course, this is a foreseeable coincidence because the mentioned velocity-dependent term coincides exactly with
$3\mathcal{H}\left(\Phi'+{\mathcal H}\Phi\right)$ owing to the linearized Einstein equation $G_{\alpha}^0=\kappa T_{\alpha}^0$.
%may serve as a possible interpretation of the physical meaning of $l$: screening is ensured by a collective hydrodynamical velocity field coupled to a collective
%gravitational potential (produced by all matter smoothed in the hydrodynamical sense).
Second, in contrast to the current paper, \citet{Hahn} did not single out the very important contribution to $\delta T_0^0$, namely, the second term in the rhs of
\rf{13}, which is directly proportional to $\Phi$ (see, however, their Appendix C, where the authors address this issue along with the connection to the approach
of \citet{Chisari}). The mentioned term is absolutely necessary for satisfying the perturbed energy conservation equation \citep{Ruslan} and leads to the
screening length $\lambda$ \rf{c5} irrespective of the velocity-dependent contribution.

\

\subsection{Yukawa Interaction and Zero Average Values}

It is important to stress that, as a manifestation of the superposition principle, the second term in \rf{40} represents the sum of Yukawa potentials
\be{c9} \phi_n=-\frac{\kappa c^2}{8\pi a}\frac{m_n}{|{\bf r}-{\bf r}_n|}\exp(-q_n)=-\frac{G_Nm_n}{c^2|{\bf R}-{\bf R}_n|}\exp\left(-\frac{|{\bf R}-{\bf
R}_n|}{\lambda}\right)\ee
coming from each single particle, with the same interaction radius $\lambda$. Such a favourable situation is possible owing to the last term in the left-hand side
(lhs) of Eq.~\rf{15}, which has been disregarded in \citep{EZcosm1} by mistake and erroneously compensated in \citep{EZcosm2} by inhomogeneous radiation of
unknown nature. Actually, such radiation must not only possess negligible average energy density (requiring additional questionable reasoning), but also exchange
the momentum with the nonrelativistic pressureless matter, despite the fact that no non-gravitational interaction between these two constituents has been assumed,
and therefore the energy-momentum interchange is strictly forbidden. Here the mentioned unpardonable omission is rectified: the ill-starred term is reinstated,
and there is no necessity in any additional interacting Universe components at all.

The sum $\sum\limits_n\phi_n$ is certainly convergent at all points except at positions of the gravitating masses, and computational obstacles do not come into
existence. In particular, the order of adding terms corresponding to different particles is arbitrary and does not depend on their locations. On the contrary,
there are certain obstacles when calculating the sum of Newtonian potentials or their gradients. Let us address the well-known formulas (8.1) and (8.3) in the
textbook by \citet{Peebles} for the gravitational potential and the peculiar acceleration, respectively, derived in the Newtonian approximation (see above):
\be{c10} \Phi\sim\int d{\bf r}'\frac{\rho|_{{\bf r}={\bf r}'}-\overline\rho}{|{\bf r}-{\bf r}'|},\quad -\nabla\Phi\sim \int d{\bf r}'\frac{\rho|_{{\bf r}={\bf
r}'}}{|{\bf r}-{\bf r}'|^3}({\bf r}-{\bf r}')\ee
up to space-independent factors being of no interest here. Substituting \rf{12} into the second integral in \rf{c10}, one gets the formula (8.5) in
\citep{Peebles}:
\be{c11} -\nabla\Phi\sim \sum\limits_n\frac{m_n}{|{\bf r}-{\bf r}_n|^3}({\bf r}-{\bf r}_n)\, .\ee
According to \citet{Peebles}, this sum is not well-defined and depends on the order of adding terms, and if one adds them in the order of increasing distances
$|{\bf r}-{\bf r}_n|$ and assumes that the distribution of particles corresponds to a spatially homogeneous and isotropic random process with the correlation
length being much less than the Hubble radius $c/H$, then this sum converges. As regards the first integral in \rf{c10}, the argumentation by \citet{Peebles}
again relies on the random process assuring convergence; however, substitution of \rf{12} splits this integral into two divergent parts: the sum of an infinite
number of the Newtonian potentials of the same sign (see also the paper by \citet{Norton} devoted to the related famous Neumann-Seeliger gravitational paradox)
and the integral of the pure Newtonian kernel.

Of course, the enumerated difficulties are absent when summing up the Yukawa-type potentials. In addition, it is interesting that in this case the particles'
distribution may be nonrandom and anisotropic. The lattice Universe model with the toroidal topology $T\times T\times T$ represents a striking example. As
explicitly demonstrated by \citet{Lattice}, in the framework of this model the gravitational potential has no definite values on the straight lines joining
identical point-like masses in neighbouring cells if the last term in the lhs of Eq.~\rf{15} is not taken into account. Evidently, the finite Yukawa interaction
range $\lambda$ arising due to this term easily resolves this challenge as well as any similar ones related to the choice of periodic boundary conditions.
Incidentally, if the space is supposed to have the usual, non-toroidal topology $R\times R\times R$, but the choice of periodic boundary conditions is made for
$N$-body simulation purposes, then the dimensions of a cell should normally be greater than $\lambda$, thereby weakening the undesirable impact of periodicity on
simulation outputs.

A noteworthy feature of the Yukawa potentials \rf{c9} consists in assuring the zero average value of the scalar perturbation $\Phi$ \rf{40}. Let us determine the
average value of a single one of them:
\be{c12} \overline\phi_n\equiv\frac{1}{V}\int\limits_Vd{\bf r}\phi_n=-\frac{\kappa c^2}{8\pi a}\frac{m_n}{V}\int\limits_V\frac{d{\bf r}}{|{\bf r}-{\bf
r}_n|}\exp\left(-\frac{a|{\bf r}-{\bf r}_n|}{\lambda}\right)=-\frac{\kappa c^2}{8\pi
a}\frac{m_n}{V}\frac{4\pi\lambda^2}{a^2}=-\frac{m_n}{V}\frac{1}{3\overline\rho}\, ,\ee
where the comoving averaging volume $V$ tends to infinity. Here the definition of $\lambda$ \rf{c5} has been used. Consequently,
\be{c13} \sum\limits_n\overline\phi_n=-\frac{1}{3\overline\rho}\cdot\frac{1}{V}\sum\limits_nm_n=-\frac{1}{3}\, ,\ee
since $(1/V)\sum\limits_nm_n\equiv\overline\rho$. Combining \rf{c13} with the first term in \rf{40}, one immediately achieves the desired result $\overline\Phi=0$
(the third term in \rf{40} is apparently zero on average in view of the different directions of particle velocities, and the same applies to the vector
perturbation ${\bf B}$ \rf{36}: $\overline{\bf B}=0$). This result means that the first-order backreaction effects are absent, as it certainly should be. Zero
average values of the first-order cosmological perturbations are expected from the very beginning, since these metric corrections are none other than linear
deviations from the unperturbed average values of the metric coefficients. Nevertheless, as shown by \citet{EZcosm2,EBV}, there exists a concrete example of the
mass distribution, which gives the nonzero average value of the gravitational potential determined by the standard prescription \rf{c10}. This problem is solved
by \citet{EBV} through introducing manually the abrupt cutoff of the gravitational interaction range with the help of the Heaviside step function. One can see now
that the same problem is strictly solved with the help of the finite Yukawa range, and the potential remains smooth together with its gradient thanks to the
smoothness of the exponential function. Obviously, the established equality $\overline\Phi=0$ takes place for an arbitrary mass distribution including that
investigated by \citet{EBV}. In addition, the well-grounded equalities $\overline{\delta T}_0^0=0$ and $\overline{\delta T}_{\alpha}^0=0$ are valid as well,
following from \rf{13} and \rf{14}, respectively.

Let us bring up and settle a related issue consisting in the following. One can easily prove that in the limiting case of the homogeneous mass distribution
$\Phi=0$ at any point, as it certainly should be. For example, on the surface of a sphere of physical radius $R$ the contributions from its inner and outer
regions combined with $1/3$ in \rf{40} give zero (see, e.g., the expression (3.12) in \citep{shell} for the gravitational potential within a spherical shell of
uniform mass density, the inner radius $R_1\rightarrow0$ and the outer radius $R_2\rightarrow+\infty$, which gives $-1/3$, with the exception of the zero mode,
which should be dropped). This means that in the considered limiting case the equation of motion of a test cosmic body reads:
\be{c14} \ddot {\bf R}=\frac{\ddot a}{a} {\bf R}\, , \ee
so the acceleration of the body is reasonably connected with the acceleration of the Universe expansion. At the same time, the described simple, but crucial test
cannot be passed by Newtonian gravity. Indeed, in the framework of the Newtonian cosmological approximation the contribution from the outer region of the
considered sphere is absent, while the contribution from its inner region generates an additional force in the rhs of Eq.~\rf{c14}, spoiling the established
correspondence between the accelerations. This demonstrates once again the superiority of the formula \rf{40} for all scales.
%
%\be{c15} \ddot {\bf R}=\frac{\ddot a}{a} {\bf R}-\frac{4\pi G_N\overline\rho}{3a^3}{\bf R}=\left(\frac{\Lambda c^2}{3}-\frac{\kappa\overline\rho
%c^4}{6a^3}\right){\bf R}-\frac{4\pi G_N\overline\rho}{3a^3}{\bf R}\, , \ee
%
%where the Friedmann equations \rf{c7} and \rf{c8} have been used to express $\ddot a/a$.

\

\subsection{Transformation of spatial coordinates}

When writing down the perturbed metric \rf{4}, the gauge choice is made in favour of the so-called Poisson/longitudinal/conformal-Newtonian gauge, by analogy with
\citet{AdamekRuth,Adamek2,Bruni}. However, it is common knowledge that there is no preferable coordinate system, so other gauges are admissible as well. The
chosen gauge is characterized, in particular, by the coincidence of the found function $\Phi$ \rf{40} with the corresponding gauge-invariant Bardeen potential
\citep{Bardeen}.
%In simple words, the scalar perturbation \rf{40} does not depend on any particular gauge.
The introduced energy-momentum fluctuations $\delta T_i^k$ also coincide with the corresponding gauge-independent quantities. For instance, let us verify that the
expression \rf{13} for $\delta T_0^0$ remains unchanged for the analogue of the so-called $N$-body gauge \citep{gauge}. This particular gauge features the
unperturbed comoving volume giving a chance of eliminating the second term in the rhs of \rf{13} and, hence, of rehabilitating the Newtonian description. In this
connection, it is necessary to show directly that this chance does not contradict the Yukawa screening of the gravitational interaction established in the Poisson
gauge. For this purpose, let us rewrite the metric \rf{4} excepting the vector perturbation ${\bf B}$:
\be{g1} ds^2=a^2\left[(1+2\Phi)d\eta^2-(1-2\Phi)\delta_{\alpha\beta}dx^{\alpha}dx^{\beta}\right]\, ,\ee
where the scale factor $a$ is a function of the conformal time $\eta$ while the scalar perturbation $\Phi$ \rf{40} is a function of $\eta$ and comoving
coordinates $x^{\alpha}$, $\alpha=1,2,3$. The transformation of coordinates
\be{g2} \eta=\tau+A,\quad x^{\alpha}=\xi^{\alpha}+\frac{\partial L}{\partial\xi^{\alpha}}\, ,\ee
where $A$ and $L$ are (first-order) functions of the new conformal time $\tau$ and new comoving coordinates $\xi^{\alpha}$, $\alpha=1,2,3$, gives
\ba{g3} ds^2&=&a^2\left[ \left(1+2\Phi+2A'+2\mathcal{H}A\right)d\tau^2+2\left(\frac{\partial A}{\partial \xi^{\alpha}}-\frac{\partial L'}{\partial
\xi^{\alpha}}\right)d\tau d\xi^{\alpha} \right.\nn\\
&-&\left. \left( (1-2\Phi+2\mathcal{H}A)\delta_{\alpha\beta}+2\frac{\partial^2L}{\partial\xi^{\alpha}\partial\xi^{\beta}} \right)d\xi^{\alpha}d\xi^{\beta}
\right]\, .\ea
Here the prime denotes the derivative with respect to $\tau$; $a$ and $\mathcal{H}$ depend on $\tau$ while $\Phi$ depends on $(\tau,\xi^{\alpha})$. Fixing $A=0$,
one immediately comes to the opportune coincidence of the fluctuations of the mixed energy-momentum tensor components with the corresponding gauge-invariant
perturbations. Despite the fact that this choice differs from that made by \citet{gauge} (where $A\neq0$), this does not affect the following main idea of the
$N$-body gauge. In accordance with the general definition \rf{11}, in the new coordinates $(\tau,\xi^{\alpha})$ instead of \rf{13} one has
\be{g4} \delta T_0^0=\frac{c^2}{a^3}\delta\rho_{\xi}+\frac{\overline\rho c^2}{a^3}\left(3\Phi-\triangle_{\xi} L\right)\, ,\ee
where $\triangle_{\xi}\equiv\delta^{\alpha\beta}\cfrac{\partial^2}{\partial \xi^{\alpha}\partial \xi^{\beta}}\,$;
\be{g5} \delta\rho_{\xi}\equiv\rho_{\xi}-\overline\rho,\quad \rho_{\xi}=\sum\limits_nm_n\delta\left(\vec{\xi}-\vec{\xi}_n\right)\, .\ee
Next, fixing $\triangle_{\xi} L=3\Phi$, one may present the energy density fluctuation \rf{g4} conformably in the form
\be{g6} \delta T_0^0=\frac{c^2}{a^3}\delta\rho_{\xi}\, .\ee
Thus, it may seem that proper use of gauge freedom ensures disappearance of the second term in the rhs of Eq.~\rf{13}. Nevertheless, the expressions \rf{13} and
\rf{g6} for $\delta T_0^0$ are equal. In order to prove this, let us use the fact that the perturbation $\delta\rho_{\xi}$ entering into \rf{g6} is not equal to
the counterpart $\delta\rho$ entering into \rf{13}. Indeed, the rest mass density $\rho$ \rf{12} is connected with $\rho_{\xi}$ \rf{g5} by means of the
relationship
\be{g7} \rho=\frac{1}{1+\triangle_{\xi}L}\rho_{\xi}\, ,\ee
where the denominator represents the Jacobian $\mathrm{det}\left(\partial x^{\alpha}/\partial \xi^{\beta}\right)$ of the comoving coordinates transformation.
Since $\rho=\overline\rho+\delta\rho$ and $\rho_{\xi}=\overline\rho+\delta\rho_{\xi}$, recalling that $L$ is the first-order quantity, from \rf{g7} one gets
\be{g8} \delta\rho_{\xi}=\delta\rho+\overline\rho\triangle_{\xi} L=\delta\rho+3\overline\rho \Phi\, .\ee
Substitution of \rf{g8} into \rf{g6} revives the gauge-independent perturbation \rf{13}. It is important to remember that positions of the gravitating masses are
described by radius-vectors which certainly depend on the choice of comoving coordinates. For instance, apparently, ${\bf r}_n\neq \vec{\xi}_n$ in the case of the
nontrivial function $L$ in \rf{g2}.

%{\bf In this connection, since the Poisson gauge is fixed from the very beginning of the present paper without spoiling orthogonality of the chosen comoving
%coordinates $x^{\alpha}$, it is logical that positions of the gravitational field sources are specified appropriately (by means of ${\bf r}_n$).}

%, the second term in the lhs of Eq.~\rf{27} and, hence, the sum of Yukawa potentials. Recapitulating, the revealed Yukawa screening is by no means a result of
%overestimation of a gauge-dependent mathematical expression. Quite the contrary, it is a fundamental gauge-invariant feature of universal gravitation.

%{\bf With the help of the nonzero function $A$, \citet{gauge} impose a comoving gauge condition along with the unperturbed comoving volume condition and arrive at
%a correction to the Newtonian geodesic equation (see their Eq.~(12)). Then the authors argue that this correction vanishes at the matter- and $\Lambda$-dominated
%stages of the Universe evolution.}

The initial displacement of particles proposed by \citet{Chisari} can be studied in the same vein. Restricting themselves to the linear relativistic perturbation
theory for large enough scales where the failure of Newtonian dynamics is expected and striving for absorption of relativistic effects into the initial conditions
for Newtonian simulation codes, the authors took advantage of the transformation of spatial coordinates
\be{g9} x^{\alpha}=\xi^{\alpha}+\delta x_{\mathrm{in}}^{\alpha},\quad \frac{\partial}{\partial \xi^{\alpha}}\left(\delta
x_{\mathrm{in}}^{\alpha}\right)=3\zeta_{\mathrm{in}}\, ,\ee
where $\zeta_{\mathrm{in}}$ stands for the initial value of the so-called comoving curvature, or curvature perturbation variable \citep{Ruth},
\be{g10} \zeta=\frac{2a\mathcal{H}\left(\Phi'+\mathcal{H}\Phi\right)}{\kappa \overline\rho c^2}+\Phi\, .\ee
Then substitution of \rf{g4}, where now $\triangle_{\xi} L$ is replaced by $3\zeta_{\mathrm{in}}$, into \rf{7} gives
\be{g11} \triangle\Phi-3{\mathcal H}(\Phi'+{\mathcal H}\Phi) = \frac{\kappa c^2}{2a}\left[\delta\rho_{\xi}+3\overline\rho
\left(\Phi-\zeta_{\mathrm{in}}\right)\right]\, . \ee
Taking into account that the introduced comoving curvature does not evolve at large scales under consideration, one can replace $\zeta_{\mathrm{in}}$ in \rf{g11}
by $\zeta$ \rf{g10}, and the subsequent cancellation of terms in the obtained equation reduces it to the following form:
\be{g12} \triangle\Phi= \frac{\kappa c^2}{2a}\delta\rho_{\xi}\, . \ee
Once again, as it follows from the first equality in \rf{g8}, $\delta\rho_{\xi}=\delta\rho+3\overline\rho\zeta_{\mathrm{in}}$. Then Eq.~\rf{g12} is reduced to its
original form before the transformation \rf{g9}, in complete agreement with the gauge invariance of the Bardeen potential.

Summarizing, there are two consistent options for cosmological simulations. On the one hand, one can resort to the initial displacement of particles
\citep{Chisari} or the $N$-body gauge \citep{gauge} and reinterpret the large-scale Newtonian $N$-body outputs as the relativistic ones. On the other hand, one
can remain faithful to the Poisson gauge and calculate the gravitational potential from the Helmholtz equation, in harmony with the reasoning by \citet{Hahn} (see
also the paper by \citet{Rampf} where the Helmholtz equation links the potential to the density perturbation at scales comparable to the horizon).

\

\subsection{Nonzero Spatial Curvature and Screening of Gravity}

The promised generalization to both cases of curved spatial geometry can be made straightforwardly. For simplicity and illustration purposes, let us restrict
ourselves to Eq.~\rf{27} and rewrite it dropping the velocity contributions (i.e. the second term in the rhs) and taking into consideration the nonzero spatial
curvature:
\be{c} \triangle\Phi+\left(3\mathcal K-\frac{3\kappa\overline\rho c^2}{2a}\right)\Phi = \frac{\kappa c^2}{2a}\delta\rho\, ,\ee
where $\mathcal K=+1$ for the spherical (closed) space and $\mathcal K=-1$ for the hyperbolic (open) space, and the Laplace operator is redefined appropriately
(see \citet{EZcosm1,Alvina}). This equation is equivalent to the equation (2.25) in \citep{Alvina} up to designations. Hence, one can make use of its solutions
derived by \citet{Alvina}, simply adjusting the notation. There seems no sense to reproduce these solutions here, but it should be emphasized that they are smooth
at any point except at particle positions (where the Newtonian limits are reached) and characterized by zero average values, similarly to the flat space case
$\mathcal K=0$.
%In addition, there is a smooth transition between the cases $\mathcal K=0$ and $\mathcal K\neq0$ in the discussed formulas.

One more important detail lies in the fact that the definition of $\lambda$ \rf{c6} remains valid not only in the curved space case, but also in the presence of
an arbitrary number of additional Universe components in the form of perfect fluids with constant or varying parameters in the equations of state like
$p=\omega\varepsilon$ (e.g., radiation with the parameter $1/3$), as one can prove \citep{Ruslan}. This hints at the universality of the presentation \rf{c6}. In
particular, the gravitational potentials derived by \citet{Alvina} may be interpreted as valid for the Universe filled with quintessence with the parameter $-1/3$
in the presence of the cosmological constant as well as the nonrelativistic pressureless matter with negligible average energy density.

Returning to the conventional cosmological model, from \rf{c6} one gets the dependence $\lambda\sim a^2$ at the radiation-dominated stage of the Universe
evolution. Since $\lambda$ may be associated with the homogeneity scale, as stated above, the asymptotic behavior $\lambda\rightarrow0$ when $a\rightarrow0$
supports the idea of the homogeneous Big Bang.

Finally, it seems almost impossible to overcome the irresistible temptation of associating the Yukawa interaction range $\lambda$ with the graviton Compton
wavelength $h/(m_gc)$, where $h$ is the Planck constant and $m_g$ is the graviton mass, in the particle physics spirit. However, one should act with proper
circumspection when discussing the massive graviton (see reasoning by \citet{Faraoni} as well as argumentation by \citet{Novello} with respect to Minkowski and de
Sitter spacetimes). It is remarkable that setting $\lambda$ equal to $\hbar/(m_gc)$, $\hbar\equiv h/(2\pi)$, gives $m_g=\hbar/(\lambda
c)\approx1.7\times10^{-33}\, \mathrm{eV}$ today (when $a=a_0$), and the ratio $1/\lambda^2=m_g^2c^2/\hbar^2$ turns out to be numerically equal to $2\Lambda/3$.
And vice versa, if one does not initially resort to the known numerical value of $\Omega_M$ and, hence, does not estimate $\lambda$ and $m_g$, the conjectural
relationship $m_g^2c^2/\hbar^2=2\Lambda/3$ (see \citet{Haranas} and Refs. therein) when $a=a_0$ may be rewritten with the help of \rf{c5} as
$9\Omega_M=4\Omega_{\Lambda}$, whence in the case of the negligible spatial curvature ($\Omega_M+\Omega_{\Lambda}=1$) one gets $\Omega_M=4/13\approx0.31$, in
solid agreement with \citet{Planck13,Planck15}. It is noteworthy as well that since $\lambda\sim a^{3/2}$ \rf{c5}, one has $m_g\sim a^{-3/2}$, so $m_g\sim 1/t$ at
the matter-dominated stage of the Universe evolution (when $a\sim t^{2/3}$). This dependence on time agrees with that found by \citet{Haranas}. At the
radiation-dominated stage $\lambda\sim a^2$, $m_g\sim a^{-2}$. Thus, $m_g\rightarrow 0$ when $a\rightarrow+\infty$ ($\Lambda$-domination prevents screening of
gravity) and formally $m_g\rightarrow+\infty$ when $a\rightarrow0$.

The established finite Yukawa range of the gravitational interaction may potentially pretend to play a key role in resolving the coincidence and cosmological
constant problems as well as developing the holographic and inflationary scenarios. Clarification and rigorous substantiation of this role overstep the limits of
the current paper.

\

\section{CONCLUSION}

The following main results have been obtained in the present paper in the framework of the concordance cosmological model:

\vspace{0.2cm}

\noindent $\bullet$ the first-order scalar \rf{40} and vector \rf{36} cosmological perturbations, produced by inhomogeneities in the discrete form of a system of
separate point-like gravitating masses, are derived in the weak gravitational field limit without any supplementary approximations (no $1/c$ series expansion, no
``dictionaries'');

\vspace{0.2cm}

\noindent $\bullet$ the obtained explicit analytical expressions \rf{36} and \rf{40} for the metric corrections are valid at all (sub-horizon and super-horizon)
scales, converge at all points except at locations of the sources (where the appropriate Newtonian limits are reached), and average to zero (no first-order
backreaction effects);

\vspace{0.2cm}

\noindent $\bullet$ both the Minkowski background limit (see \rf{c2} and \rf{c3}) and the Newtonian cosmological approximation (see \rf{c4}), which is widely used
in modern $N$-body simulations, represent particular limiting cases of the constructed scheme and serve as its corroboration;

\vspace{0.2cm}

\noindent $\bullet$ the velocity-independent part of the scalar perturbation \rf{40} contains a sum of Yukawa potentials produced by inhomogeneities with the same
finite time-dependent Yukawa interaction range \rf{c5}, which may be connected with the scale of homogeneity, thereby explaining the existence of the largest
cosmic structures;

\vspace{0.2cm}

\noindent $\bullet$ the general Yukawa range definition \rf{c6} is given for various extensions of the $\Lambda$CDM model (nonzero spatial curvature, additional
perfect fluids), and advantages of the established gravity screening are briefly discussed.

\vspace{0.2cm}

Based on the obtained results, it should be not too difficult to construct similarly an appropriate scheme for the second-order cosmological perturbations
including the tensor ones. Accomplishment of this quite possible technical mission would predict, in particular, the backreaction effects. It is expected that the
second-order metric corrections will be much smaller than the first-order ones at arbitrary scales. Besides, the direct generalization of the elaborated approach
to the case of alternative (nonconventional) cosmological models, for example, those replacing the $\Lambda$-term by some other dynamical physical substance,
serving as dark energy and also fitting all data, is straightforward and can be made with hardly any trouble. Then, simulating nonlinear dynamics at arbitrary
scales, predicting formation and evolution of large cosmic structures, determining the influence of metric corrections on propagation of photons through the
simulation volume, etc, one can actually probe cosmology and potentially distinguish among different competing representations of the dark sector. Of course,
extra effort and care are required for constituting a link between physical quantities extracted from relativistic simulations and observables measured in galaxy
surveys such as redshifts and positions in the sky (see, e.g., \citet{Bonvin,Yoo}).

Thus, the developed cosmological perturbations theory covering the whole space, in combination with such future high-precision surveys as Euclid \citep{Euclid},
approaching the Hubble horizon scale, may essentially deepen our knowledge about the amazing world we live in.

%varying gravitational constant as a tool for compensating varying cosmological constant;
%other possibilities include non-gravitational interactions or change of the vacuum EoS, corresponding Refs.

\

\section*{ACKNOWLEDGEMENTS}

This work was supported by NSF CREST award HRD-1345219 and NASA grant NNX09AV07A. I would like to thank the anonymous Referee for valuable comments which have
considerably improved the discussion of the derived results. I am also grateful to my colleague Prof. Diane~Markoff for the careful review of their presentation.

\

\end{document}